\documentclass[preprint2]{aastex}

\def\lsim{\mathrel{\rlap{\lower 4pt \hbox{\hskip 1pt $\sim$}}\raise 1pt \hbox {$<$}}}
\def\gsim{\mathrel{\rlap{\lower 4pt \hbox{\hskip 1pt $\sim$}}\raise 1pt \hbox {$>$}}}

\newcommand{\msun}{M_{\odot}}

\shorttitle{Hypernova Light Curves} 
\shortauthors{Maeda et al.}

\begin{document}

\title{A Two-Component Model for the Light Curves of Hypernovae}

\author{Keiichi Maeda\altaffilmark{1}, Paolo A. Mazzali\altaffilmark{1,2},
Jingsong Deng\altaffilmark{1,3}, Ken'ichi Nomoto\altaffilmark{1,3}, 
Yuzuru Yoshii\altaffilmark{3,4}, Hiroyuki Tomita\altaffilmark{1}, 
and Yukiyasu Kobayashi\altaffilmark{5}
}

\altaffiltext{1}{Department of Astronomy, School of Science, 
University of Tokyo, 7-3-1 Hongo, Bunkyo-ku, Tokyo 113-0033, Japan: 
maeda@astron.s.u-tokyo.ac.jp}
\altaffiltext{2}{INAF - Osservatorio Astronomico, 
Via Tiepolo, 11, 34131 Trieste, Italy}
\altaffiltext{3}{Research Center for the Early Universe, School of Science, 
University of Tokyo, Hongo 7-3-1, Bunkyo-ku, Tokyo 113-0033, Japan} 
\altaffiltext{4}{Institute of Astronomy, School of Science, University of Tokyo, 
2-21-1 Osawa, Mitaka, Tokyo 181-0015, Japan}
\altaffiltext{5}{National Astronomical Observatory, 
2-21-1 Osawa, Mitaka, Tokyo 181-8588, Japan}

\begin{abstract}
The light curves of 'hypernovae', i.e. very energetic supernovae 
with $E_{51} \equiv E/10^{51}$ergs $\gsim 5-10$ are characterized 
at epochs of a few months by a phase of linear decline. 
Classical, one-dimensional explosion models fail to simultaneously
reproduce the light curve near peak and at the linear decline phase.
The evolution of these light curves may however be explained by a simple  
model consisting of two concentric components. The outer component
is responsible for the early part of the light curve and for 
the broad absorption features observed in the early spectra of
hypernovae, similar to the one-dimensional models. In addition, a
very dense inner component is added, which reproduces the linear
decline phase in the observed magnitude-versus-time relation for 
SNe 1998bw, 1997ef, and 2002ap. 
This simple approach does contain one of the main features of jet-driven, 
asymmetric explosion models, namely the presence of a dense core. 
Although the total masses and energies derived with the two-component model 
are similar to those obtained in previous studies which also adopted 
spherical symmetry, this study suggests that the ejecta 
are aspherical, and thus the real energies and masses may deviate 
from those derived assuming spherical symmetry. 
The supernovae which were modeled are divided into two groups,
according to the prominence of the inner component: the inner
component of SN 1997ef is denser and more $^{56}$Ni-rich, relative
to the outer component, than the corresponding inner components of
SNe 1998bw and 2002ap. These latter objects have a similar
inner-to-outer component ratio, although they have very different 
global values of mass and energy. 
\end{abstract}


\section{Introduction}

Triggered by the discovery of the unusual supernova SN~1998bw in
the error box of the enigmatic $\gamma$-ray burst, GRB~980425
(Galama et al. 1998; Iwamoto et al. 1998), a new class of
core-collapse supernovae, similar to SN~1998bw, has been
suggested. They show very broad absorption features in their
early spectra, which indicates the presence of a large amount of 
material moving at high velocities (see Nomoto et al. 2001 for 
a review).

Among the best-studied hypernova candidates are SN~1998bw 
(Iwamoto et al. 1998; Woosley, Eastman, \& Schmidt, 1999; 
Nakamura et al. 2001; Mazzali et al. 2001),
SN~1997ef (Iwamoto et al. 2000; Mazzali, Iwamoto, \& Nomoto 2000), and
SN~2002ap (Mazzali et al. 2002). All these supernovae (SNe) were 
spectroscopically classified as Type Ic SNe (SNe~Ic). 

Because of their spectral characteristics, they were all modeled
as core collapse-induced explosions of bare carbon-oxygen stars. 
The properties of the individual SNe vary, but they all have in 
common a large explosion kinetic energy,
$E_{51} \equiv E/10^{51}$ergs$\gsim 5-10$, which is much 
larger than in normal SNe. Accordingly, these objects were 
termed ``hypernovae'' (e.g. Iwamoto et al. 1998). 
Some of the model parameters used to reproduce the properties 
of these SNe near peak are listed in Table 1.

Actually, the term ``hypernova'' was first introduced by Paczynski (1998) 
to describe the entire GRB/afterglow event, based on the very energetic nature 
of the GRB phenomenon. He linked it to the cataclysmic death of massive stars 
and the formation of black holes, as in the failed-supernova model of Woosley 
(1993). 
Though these studies focused on modeling GRBs, SN~1998bw/GRB~980425 
demonstrated that such a scenario may also be applied to highly energetic 
SN explosions. The term ``hypernova'' has meanwhile been 
used to describe such an event (e.g., Iwamoto et al. 1998). 
We follow this terminology in this paper.

The mechanism causing such extremely energetic explosions is however 
not yet agreed upon. Analyzing the light curve and spectra at 
intermediate (i.e. between the photospheric and nebular phase, 
$\sim 50-200$ days) and late phases (nebular spectra, 
$\gsim 200$ days) must help answering this question. 
At increasingly advanced phases, in fact, the inner part of the 
ejecta dominates the optical output of the SN, thus providing
hints on the properties of the explosion. 

The spherical hydrodynamical models used to fit the early 
phases of SNe 1998bw (Nakamura et al. 2001), 
1997ef (Iwamoto et al. 2000; Mazzali et al. 2000), 
and 2002ap (Mazzali et al. 2002; Yoshii et al. 2003) do not  
reproduce well the later phase light curves. 
Specifically, they fail to reproduce the linear decline 
in magnitude which is observed in all hypernovae at phases of 
about one to a few 100 days (McKenzie \& Schaefer 1999). 
The observed slopes are $\sim 0.018$ mag day$^{-1}$ for 
SNe 1998bw and 2002ap, and $\sim 0.01$ mag day$^{-1}$ for SN 1997ef. 
The latter is in good agreement with the slope expected for full 
trapping of the $\gamma$-rays emitted by the radioactive decay of 
$^{56}$Co. 
All models predict late-phase optical luminosities much fainter 
than observed. 

A similar problem may have been encountered for some SNe IIb/Ib/Ic. Clocchiatti \& 
Wheeler (1997) pointed out that the V light curves of SN Ib 1983N, SN Ic 1983V, 
SN IIb 1993J, and a few others are very similar to one another, showing linear decline 
after $\sim 50$ days (see also Clocchiatti et al. 1996; Clocchiatti et al. 1997). 
Those authors further showed that 
this slope is naturally explained as a consequence of the asymptotic behavior 
of the $\gamma$-ray deposition function which becomes independent from the 
model parameter involved, e.g., $M_{\rm ej}$, $E$. They suggested that the inner 
ejecta may not expand homologously. However, this problem was raised by comparing 
the theoretical bolometric light curve with observed V light curves, in contrast to 
the case for SNe 1998bw, 1997ef, and 2002ap. 
We note that the 
detailed light curve calculations based on spherically symmetric hydrodynamical 
models by Iwamoto et al. (1997)  (see their figure 11) 
and Blinnikov et al. (1998) for SN 1993J gave reasonably good fits to the observed bolometric light curve from near the shock breakout up to $t \sim 300$ and 120 days, respectively.

In this paper, we concentrate on the light curves of the hypernovae 
SNe 1997ef, 1998bw, and 2002ap. 
The same scenario used to model those light curves may possibly apply to the 
other SNe mentioned above. 
We first clarify why the original hydrodynamic models fail to fit the 
light curves at advanced epochs, then we suggest an alternative 
approach, which can yield good fits to the light curves of 
SNe~1998bw, 1997ef, and 2002ap at both early and late phases. 
The model we present consists of two components in density. 
An outer, high velocity component is responsible for the early 
phase observations, while an inner, dense, low velocity component 
is responsible for the late phase. At intermediate epochs, these 
two components have different optical depths to $\gamma$-rays, and 
this is the central point of this model.

The suggestion that a low-velocity, high-density central zone may 
exist in hypernovae, which may explain the behavior of the light 
curve was made for SN~1997ef by Iwamoto et al. (2000) and by 
Mazzali et al. (2000), who also found spectroscopic support for this
hypothesis in the unexpectedly long duration of the photospheric phase, 
and for SN~1998bw by Mazzali et al. (2001) based on the presence of a 
narrow component in the profile of the nebular line of [OI] 6300\AA. 
Nomoto et al. (2001) and Nakamura et al. (2001) then showed that, while 
the early light curve of SN~1998bw requires a high-energy model, the 
late-time light curve is well reproduced by a lower energy model with 
a smaller $^{56}$Ni mass, which is however too dim at early phases. 
H\"oflich, Wheeler, \& Wang (1999) also suggested that a modification 
of the structure of the inner ejecta will remedy the fact that the light curves 
of their aspherical models decline slightly too steep after maximum. 
However what concerned them are the light curves before day $40$, 
and we are studying the phases $> 50$ days.

In Section 2, the usefulness of a two-component model 
through its effect on $\gamma$-ray deposition in the SN ejecta 
is shown using a simple parameterized model. 
More accurate models are then constructed for the three hypernovae.
These models are still within the context of spherically symmetric explosion, 
but they do take spectral observations into account (Section 3). 
Using this approach, a possible physical link among the various 
hypernovae is revealed which is more profound than the differences among 
the various objects we have studied. This link may hold the key to 
understanding the explosion mechanism of hypernovae, and possibly of some
normal SNe Ibc as well.

\section{Simple $\gamma$-ray Deposition Models}
		
To set the scene, we computed bolometric light curves for the three 
hypernovae using the Monte Carlo light curve code described in Section 3. 
We used models CO138 of Nakamura et al. (2001) for SN~1998bw, CO100 
of Mazzali et al. (2000) for SN~1997ef, and the re-scaled version of 
CO100 used by Mazzali et al. (2002) for SN~2002ap.
The properties of these models are summarized in Table 1.

The resulting synthetic light curves are compared to the observed 
hypernova light curves in Figure 1. Observed bolometric points are 
from Patat et al. (2001) for SN~1998bw and from Mazzali et al. (2000) 
for SN~1997ef. 
The bolometric light curve of SN~2002ap was 
constructed using mainly $UVB(RI)_cJHK$ photometry obtained with the 
MAGNUM telescope (Yoshii et al. 2003; Y. Yoshii et al., in preparation),  
combined with the data obtained at Wise Observatory (Gal-Yam et al. 2002) 
and at State Observatory, India (Pandey et al. 2003). 

As shown in Figure 1, the original models fit the light curves around 
peak very well. At phases $\gsim 50$ days, however, they fail to 
reproduce the observed linear decline in the magnitude versus time 
relation: the synthetic curves decline much faster than observed, as 
already mentioned in previous studies (McKenzie \& Schaefer 1999; 
Nakamura et al. 2001; Mazzali et al. 2000).

In order to give a simple explanation of the reason for this failure, 
we investigate the issue of $\gamma$-ray transport in supernova ejecta 
using a simple model. 
A similar approach was used by e.g. Clocchiatti \& Wheeler (1997). 
In SN ejecta, $\gamma$-rays and positrons are generated through 
the decay chain $^{56}$Ni $\rightarrow$ $^{56}$Co $\rightarrow$ $^{56}$Fe. 
A fraction of the $\gamma$-rays are thermalized and deposit their energy  
in the ejecta. This thermal energy is the observed optical output of a 
supernova (e.g. Axelrod 1980). 
The energy deposited in SN ejecta can be estimated using the formula 
\begin{equation}
  L_{\rm opt} = M(^{56}{\rm Ni}) e^{-t_{\rm d}/113} \left[ \epsilon_{\gamma} 
(1 - e^{-\tau}) + \epsilon_{e^{+}} \right] , 
\end{equation}
where the optical depth to $\gamma$-rays decreases with time and is expressed as 
\begin{equation}
    \tau  =  \kappa_{\gamma} \rho R = \tau_{0} {t_{\rm d}}^{-2} .
\end{equation}
Here the decay time of $^{56}$Co is 113 days, 
$t_{\rm d} \equiv t/1 {\rm day}$, and $\tau_{0}$ is 
the optical depth to $\gamma$-rays at $t_{\rm d} = 1$.
In equation (1), $M$($^{56}$Ni) is the mass of $^{56}$Ni 
before the onset of radioactive decay. 
The energy inputs by $\gamma$-rays and positrons are given as 
$\epsilon_{\gamma} = 6.8 \times 10^9$ erg s$^{-1}$ g$^{-1}$ 
and $\epsilon_{e^{+}} = 2.4 \times 10^8$ erg s$^{-1}$ g$^{-1}$,  
respectively. 
Positrons are assumed to be 
fully trapped {\it in situ } (Axelrod 1980). 
We set $\kappa_{\gamma}$, the effective $\gamma$-ray opacity, to
$0.027$ cm$^{2}$g$^{-1}$, which is reasonable in the gray atmosphere 
problem for $\gamma$-ray transport in SN ejecta (Axelrod 1980; 
Sutherland \& Wheeler 1984).

Equations (1) and (2) yield an approximate light curve 
at late phases i.e., when the transfer of optical photons 
becomes insignificant, 
if an appropriately averaged value is used for $\tau_{0}$. 
The value of $\tau_{0}$ depends of course on the combination of mass 
and energy, but it is also obviously affected by the geometry. 

For a homogeneous sphere, $\tau_{0}$ can be expressed as 
\begin{equation}
    \tau_{0}  \sim  1000 \times \frac{(M/\msun)^{2}}{E_{51}} .
\end{equation}
Hydrodynamical simulations of supernova explosions in one dimension usually 
yield a density structure consisting of an approximately constant-density 
central part and a power-law outer part (e.g., Nakamura et al. 2001). 
The central part dominates the optical depth to $\gamma$-rays, and so 
it is in this region that most of the energy potentially available for 
optical output is deposited. 
Thus, the ejecta can be described to a first approximation as 
a homogeneous sphere if the explosion is spherically symmetric. 
Equation (3) can be easily generalized to allow for a radial 
density gradient (Clocchiatti \& Wheeler, 1997). 

Nature, however, may much be more complicated than this idealized situation. 
It has been suggested that Rayleigh-Taylor instabilities and/or jet-induced 
explosions may cause a deviation from spherical symmetry. 
Though Rayleigh-Taylor instability-induced mixing introduces local density 
fluctuations in the ejecta (e.g., Kifonidis et al. 2000), 
the ejecta are basically spherical on the large scale. 
In addition, it is unlikely that significant mixing by Rayleigh-Taylor 
instability occurs in a bare C+O star (e.g., Shigeyama et al. 1990). 
Thus Equation (3) could give a rough estimate of $\tau_0$. 
However, a jet-induced explosion probably changes the estimate of $\tau_0$ 
significantly. A naive estimate of the lower limits of $M$ and $E$ may be 
obtained assuming that the ejecta are a homogeneous bipolar cone with opening 
half-angle $\theta$. In this case,  
\begin{equation}
   \tau_{0}  \sim  1000 \times \frac{(M/\msun)^{2}}{E_{51}} 
   \times (1 - \cos\theta)^{-1} ,
\end{equation}
as $M$ and $E$ are scaled as $(1-\cos\theta)^{-1}$. 

In any case, the precise determination of $M$ and $E$ is beyond the scope 
of this study. 
In this section, we proceed with the analysis in terms of $\tau_0$. 
The estimate of $M$ and $E$ is given in the next section assuming 
spherical symmetry, and this may be used to analyze the light curve 
based on aspherical geometry, which we postpone to the future. 

After the peak, the effect of the optical opacity in delaying photon diffusion 
becomes smaller and smaller, and the optical luminosity $L_{\rm opt}$ 
approaches the deposited $\gamma$-ray luminosity. 
The decline rate of the light curve is thus 
\begin{eqnarray}
 \frac{dM_{\rm Bol}}{dt_{\rm d}} 
& = & -2.5 \frac{d\log L_{\rm opt}}{dt_{\rm d}} \nonumber \\
 \sim  2.2  & \times & \left[ \frac{1}{226} 
+ \frac{1}{t_{\rm d}} \left\{ \tau_{0} {t_{\rm d}}^{-2} 
\frac{\exp(-\tau_{0} {t_{\rm d}}^{-2})}
{1-\exp(-\tau_{0} {t_{\rm d}}^{-2})} \right\} \right] 
\nonumber \\ 
& & \qquad \qquad \qquad \qquad {\rm mag \ day}^{-1}, 
\nonumber \\
=  2.2  & \times &
\left[ \frac{1}{226} + \frac{1}{t_{\rm d}} 
\left\{ 1 - \frac{1}{2} 
\frac{\tau_{0}}{{t_{\rm d}}^2} 
+ O \left( \left( \frac{\tau_{0}}{{t_{\rm d}}^2} \right)^2 \right) \right\} \right] 
\nonumber \\ 
& & \qquad \qquad \qquad \qquad {\rm mag \ day}^{-1}. 
\end{eqnarray}
Here we neglect $L_{e^{+}}$. 
This approximation holds as long as 
$L_{\gamma} (1 - \exp(-\tau)) \sim L_{\gamma} \tau >> L_{e^{+}}$, 
thus $t_{\rm d} \lsim t_{e^{+}} \equiv 5.3 \sqrt{\tau_0}$. 

As noted by Clocchiatti \& Wheeler (1997), as long as $\tau \ll 1$ 
(which corresponds to $t_{\rm d} \gsim t_{\gamma} \equiv \sqrt{\tau_{0}}$, 
given the strong dependence of $\tau$ on $t_{\rm d}$),
the decline rate depends only on $t_{\rm d}$ and is essentially independent 
of any physical properties of the ejecta as given by the expression
\begin{eqnarray}
   \frac{dM_{\rm Bol}}{dt} \sim 2.2 \times
\left( \frac{1}{226} + \frac{1}{t_{\rm d}} \right) \quad {\rm mag \ day}^{-1} ,
\nonumber \\
\quad t_{\gamma} \lsim t_{\rm d} \lsim t_{e^{+}} .
\end{eqnarray}
The decline rates predicted by equation (6) are 
$0.054$, $0.032$, $0.024$, and $0.021$ mags day$^{-1}$ at 
$t_{\rm d} = 50$, $100$, $150$, and $200$, respectively. 
These rates shown in Figure 1 are good approximations 
at the appropriate phases for the model curves listed in Table 1.
All these rates are significantly larger than the simple $^{56}$Co decline rate, 
0.01 mag day$^{-1}$, which is often used in parameterized light curve studies. 
This is because at these phases the light curve is dominated by 
$\gamma$-ray deposition, which is however decreasing with time. 

We have assumed that positrons are fully and instantaneously trapped 
in the SN ejecta. However, it has been suggested, both observationally
(Cappellaro et al. 1997) and theoretically (Milne, The, \& Leising 2001) 
that positrons can escape the ejecta of Type Ia SNe at very late epochs 
($t_{\rm d} >200 \sim t_{e^{+}}$). 
Though a detailed study of positron transport in SN Ib/c does not exist, 
the effect of positron escape can be neglected at epochs when 
$t_{\rm d} \lsim t_{e^{+}}$, such as those relevant to this study, as 
$\gamma$-rays still dominate then. 
Thus, the details of positron transport do not significantly 
affect the results of the current study. 

For models CO138 and CO100 rescaled, used respectively for SNe 1998bw 
and 2002ap, $t_{\gamma} = 48$ and $33$. 
Thus, the decline rate of the synthetic light curves obtained from these  
models changes significantly in the period 50--200 days, and especially 
during the first part of this period, since equation (6) predicts a 
smaller change in the rate at more advanced epochs. 
In contrast, observations over this same period show a linear decline 
with an almost constant rate of $\sim 0.018$ mag day$^{-1}$ for both SNe.

For model CO100, equation (6) applies for $t_{\rm d} \gsim t_{\gamma} = 66$. 
At $t_{\rm d} < t_{\gamma}$, equation (5) predicts a smaller decline rate for 
larger $\tau_{0}$. 
In addition, a small temporal evolution of the rate is expected for large 
$\tau_{0}$, as the terms proportional to ${t_{\rm d}}^{-1}$ and to 
$-\tau_{0} {t_{\rm d}}^{-3}$ compete with each other. The value of 
$\tau_{0}$ in CO100 is significantly larger than in CO138 and CO100 rescaled, 
and therefore this model does not show the significant change of the 
decline rate seen in models CO138 and CO100 rescaled around $\sim 50$ days. 
For $t_{\rm d} \gsim 66$, the decline rate changes only marginally in   
accordance with equation (5) (here the contribution from the $\tau_{0}$ 
term is not negligible), and at advanced epochs it approaches equation 
(6), which also predicts a small temporal evolution of the rate.  
Therefore the light curve for CO100 declines almost linearly at all 
epochs except around the peak. 
The rate of this decline, however, is twice the observed one of SN~1997ef, 
which is very close to the case of full $\gamma$-ray trapping.

According to the above arguments, one can speculate that the original 
models fail to fit the late-time light curves because their 
$\gamma$-ray optical depths is too small, so that the synthetic curves 
drop below the full $\gamma$-ray trapping line too early. 

To fit the linear decline observed at $50-200$ days, $t_{\gamma} = 100-200$ 
for SNe 1998bw and 2002ap, and $t_{\gamma} > 200$ for SN1997ef to allow 
for full $\gamma$-ray trapping. This requires dense ejecta, with large 
optical depth to $\gamma$-rays. This could be achieved if the velocity of
the ejecta were smaller. However, such dense ejecta would probably not 
be reconciled with the early phase light curve, which could be reproduced 
by the original models with a smaller optical depth, or with the spectra, 
which required high ejecta velocities. 

The only way which appears reasonable to overcome this contradiction in a 1D
model is to regard the ejecta as composed of an outer region with small optical 
depth - to fit the early phase - and of an inner region with large 
optical depth to reproduce the late phase observations. 
Actually, this cannot physically be achieved in a spherically symmetric
explosion model, as such models predict too large velocities in the inner 
regions for large $E$ or too small velocities in the outer regions for small 
$E$. Thus ultimately aspherical geometries must be considered. 
However, the basic point can be demonstrated in principle using a 1D model 
and the parameterized formulation developed above as this is basically 
independent of the assumed geometry in terms of $\tau_0$. 

Let us think of the case where the ejecta consist of two shells (or clumps) 
with different optical depths to $\gamma$-rays.
The energy deposited in the ejecta by $\gamma$-rays and positrons is
roughly expressed as follows:
\begin{eqnarray}
L_{\rm opt} =  
M(^{56}{\rm Ni})_{\rm in} e^{-t_{\rm d}/113} \left[ \epsilon_{\gamma} 
(1 - e^{-\tau_{\rm in}})  +  \epsilon_{e^{+}} \right] \nonumber \\
+  M(^{56}{\rm Ni})_{\rm out} e^{-t_{\rm d}/113} 
\left[ \epsilon_{\gamma} (1 - e^{-\tau_{\rm out}})  +  \epsilon_{e^{+}} \right] .
\end{eqnarray}
The variables have the same meaning as in equation (1), with
subscripts referring to the inner and the outer components.
The backward heating of the inner part by $\gamma$-rays
emitted in the outer component is neglected.
The optical depth decreases as the ejecta expand:
\begin{eqnarray}
\tau_{\rm in} & = & \kappa_{\gamma} \rho_{\rm in} R_{\rm in} 
= \tau_{{\rm in},0} {t_{\rm d}}^{-2} ,\\ 
\tau_{\rm out} & = & \kappa_{\gamma} \rho_{\rm out} (R_{\rm out}-R_{\rm in}) 
= \tau_{{\rm out},0} {t_{\rm d}}^{-2} .
\end{eqnarray}
Thus, $L_{\rm opt}$ can be estimated to a first approximation as a function 
of $M$($^{56}$Ni)$_{\rm i}$ and $\tau_{{\rm i},0}$. 

Good fits to the observed light curves can be obtained from equation (7) as
shown in Figure 2, where we use the parameters listed in Table 2.
The two-component model presented here significantly improves the fits to 
the observed bolometric light curves if appropriate parameters are selected.
In particular, we adopted $t_{{\rm in},\gamma} \sim 160$ for SN~1998bw 
and $t_{{\rm in},\gamma} \sim 130$ for SN~2002ap.
At a time $t_{\rm d} = t_{{\rm in},\gamma}$, the inner component already dominates
$\gamma$-ray deposition, producing thereafter a linear decline with the rate
predicted by equation (6).
In the intermediate phase, between the peak and $t_{{\rm in},\gamma}$, 
the contributions of the two components are comparable, with the 
contribution from the inner component very nearly following the Co 
decay line as this component is optically thick to $\gamma$-rays. 
The combination of the two components again results in an almost 
linear behavior.
For SN~1997ef, the observed slow decline between days $50 - 150$ requires
$t_{{\rm in},\gamma} \gsim 200$, setting a lower limit 
$\tau_{{\rm in},0}/1000 \gsim 40$. 
Because the observed slope ($\sim 0.01$ mag day$^{-1}$) is identical 
to that expected by full trapping of $\gamma$-rays, only the lower limit 
is determined for SN 1997ef. 
(However, too dense ejecta will cause the light curve shape around the peak 
too broad by delaying the appearance of optical photons. This effect sets 
the upper limit of $\tau_{{\rm in},\gamma}/1000 \lsim 70$. See the 
next section.)

To fit the light curves, we introduced four parameters, i.e., 
$M$($^{56}$Ni)$_{\rm i}$ and $\tau_{{\rm i},0}$ in Equation (7), 
which has two more parameters than in Equations (1) and (2). 
Fortunately, these parameters can be derived rather 
uniquely to fit the light curves. 
Figure 3 shows the dependence of the theoretical light curve on 
(a) $\tau_{{\rm in}, 0}$, (b) $\tau_{{\rm out}, 0}$, and (c) 
the ratio $M$($^{56}$Ni)$_{\rm in}/M$($^{56}$Ni)$_{\rm out}$. 
Since the different parameters affect the light curve shape 
in different ways, we
constrain the parameters for each object using the following strategy:  
(1) The value of $\tau_{{\rm in}, 0}$ determines the slope 
at $t_{\rm d} \gsim 50$. 
(2) The value of $\tau_{{\rm out}, 0}$ determines the epoch of the 
transition to the liner behavior and the slope at $t_{\rm d} \lsim 50$. 
(3) The ratio $M$($^{56}$Ni)$_{\rm in}/M$($^{56}$Ni)$_{\rm out}$ 
determines the contrast between peak and tail luminosity. 
(4) Finally, the total $^{56}$Ni mass scales the overall brightness. 

Obviously, these parameterized values of $\tau_{{\rm i},0}$ are the result of 
adopting an appropriate structure of the ejecta. 
For example, if the ejecta are represented as a homogeneous sphere, the 
combinations of $M_{\rm i}$ and $E_{\rm i}$ that yield a particular 
value of $\tau_{{\rm i},0}$ can be obtained from Equation (3). 
Analyzing light curves in general does not yield $M$ and $E$ independently 
(e.g., Iwamoto et al. 1998). 
In the next section we discuss how $M_{\rm i}$ and $E_{\rm i}$ 
can be meaningfully chosen starting from 
the original one-dimensional explosion models.

\section{Light Curve Models}

In order to confirm the results of the previous section, we performed 
light curve calculations using the Monte Carlo light curve code 
described in Cappellaro et al (1997).
The code treats the transport not only of $\gamma$-rays but also of 
optical photons as a random walk problem, taking into account the time 
delay between the emission of a photon following the deposition of a 
$\gamma$-ray and its escape from the SN ejecta. 
Thus it can compute the light curve at all epochs, from very early 
to late phases.

The code adopts the gray approximation, with constant opacities
$\kappa_{\gamma} = 0.027$ cm$^{2}$g$^{-1}$ (see Section 2)
and $\kappa_{\rm opt} = 0.06$ cm$^{2}$g$^{-1}$.
Although in reality $\kappa_{\rm opt}$ changes with time and 
position in the ejecta, as it depends on composition and temperature,
Mazzali et al. (2000) found that for a reasonable choice of a constant
$\kappa_{\rm opt}$ the code can reproduce the early phase light curve 
of SN~1997ef.
Actually, we find that the early phase light curves of all three 
hypernovae are reproduced very well by the synthetic light curves 
obtained using the original value $\kappa_{\rm opt} = 0.06$ cm$^{2}$g$^{-1}$ 
and the respective original hydrodynamical models listed in Table 1. 
The fits are shown in Figure 1.
Since these models also provide good fits when a more sophisticated 
radiation transport code with varying $\kappa_{\rm opt}$ is used 
(e.g. Nakamura et al. 2001), we can be confident that the selected value 
of $\kappa_{\rm opt}$ is appropriate, at least in the early phase.

The characteristics of the inner and outer components of the ejecta 
of the various hypernovae must now be defined so that a good fit 
to their light curves can be obtained. 
As for the outer region, the model should be constructed in such a way that the 
density distribution in the region where the early-time spectra are formed 
($v \gsim 10000$ km s$^{-1}$) is similar to that of the original models 
CO138 (SN~1998bw), CO100 (SN~1997ef), and re-scaled CO100 (SN~2002ap), 
as these can reproduce the evolution of the photospheric velocities. 
For the inner component, in order to keep the number of free parameters 
to a minimum, we use a homogeneous sphere which extends to $v_{\rm in}$,  
the inner velocity of the outer component.
An indication of the value of $v_{\rm in}$ can be obtained from the 
widths of the nebular lines, since it is probably the dense inner region  
that mostly contributes to these features at late phases. 
In each component, $^{56}$Ni is distributed virtually homogeneously. 

It must be noted that the original models have inner cutoffs at rather large
velocities. There is however evidence that the real ejecta are also distributed
at lower velocities, as shown by the advanced-epoch spectra of SN~1997ef (Mazzali
et al. 2000) and by the narrow nebular lines of both SN~1998bw (Mazzali et al.
2001) and SN~2002ap. In the two-component models presented here this inner 
region is filled completely. 

The masses of the two components are chosen so that the observed light curves
can be reproduced. 
The mass of the outer component is varied with respect to the original models 
CO138, CO100, and rescaled CO100, but the energy contained is unchanged.  
This is done by reducing the mass mostly at the lower velocities 
of the outer components, leaving the density distribution at 
$v \gsim 10000$ km s$^{-1}$ as in the original models. 
This preserves the behavior of the photospheric velocity in the early phases,
and at the same time it allows the inner regions to be filled without
significantly affecting the diffusion time of the photons, and hence the 
light curves at the early phases. 
The mass of the inner component is then determined by changing the density 
in the inner region to fit the entire evolution of the light curves. 

The parameters we used to define the two components are listed in Table 3, 
and the synthetic light curves are shown in Figure 4.
We note that the results of the previous section suffer from the well-known
degeneracy in the estimate of mass and energy (e.g. Iwamoto et al. 1998).
However, our two-component models are built using information from both the
early-time spectra (preserving the original hydro models - albeit with the
appropriate rescaling - to describe the outer components), and from the 
late-time, nebular spectra (setting the boundary between the two components, 
$v_{\rm in}$, in a manner consistent with the observed line widths).
This allows us to define a unique set of the values $M_{\rm ej}$, $E_{51}$, 
and $M$($^{56}$Ni) for both the inner and the outer components.

For SNe~1998bw and 2002ap, we can reproduce the observed light curves 
during their entire evolution, except for the first point. 
We are not concerned by the slight disagreement for the first point. 
As is well known, the light curve at such an early phase is sensitive 
to the details of the ejecta structure and Ni distribution near the 
surface, which are not subjects of this study. 
The outer, high-velocity material determines the shape of the 
light curve around peak, while the inner, dense part produces 
the linear decline in the intermediate phase. 

For SN~1997ef the fit is marginal. A very dense inner component is required 
to slow down the propagation of the optical photons for this supernova. 
This makes the peak very broad, but it smoothes out the sharp transition 
from the peak to the linear behavior which should occur around day 100 
(see previous section). This was successfully - but incorrectly - 
reproduced by the simple deposition model (Figure 2) because optical 
photon transfer was neglected there. 
We note, however, that $\kappa_{\rm opt}$ is of larger importance in 
SN~1997ef than in SNe~1998bw and 2002ap: because of the denser inner 
component in SN~1997ef, $\kappa_{\rm opt}$ influences the propagation of photons
for a significantly longer time here, and thus the light curve is sensitive to
changes in $\kappa_{\rm opt}$ occurring on a timescale of $\sim 100$ days. 
Thus, the assumption of constant $\kappa_{\rm opt}$ may not be as good an
approximation in SN1997ef as in the other two hypernovae.
More realistic radiation calculations may yield a better fit for SN~1997ef: 
if $\kappa_{\rm opt}$ decreases with time the transition between the peak 
and the linear decline phase would be sharper.

\section{Discussion}

In this paper, we presented a model which can explain within a single 
qualitative scenario the entire evolution of the light curves of all three 
hypernovae well observed so far.
The model assumes that two components of the ejecta can be defined in every
hypernova: an outer, high-velocity component and an inner dense component.
The outer component determines the light curve around the peak, but its effect
fades rapidly. After peak, the inner component dominates and reproduces
the observed linear decline of the light curves at intermediate phases.
At very late phases, only the positrons contribute to the light curve.
Since positrons are assumed to deposit fully, the brightness of the light 
curves at such late phases depends only on the total $^{56}$Ni mass in the 
ejecta.

The total masses and energies resulting from the sum of the inner and the 
outer components do not significantly differ from the corresponding values 
derived by previous studies of the early phases (listed in Table 1), 
and neither does the estimate of the total $^{56}$Ni mass. 
Actually, the Ni masses derived with the LC model are somewhat larger than the
original ones. This increased Ni mass is mostly the result of adding the inner
component (partially balanced by some redistribution of the Ni towards the inner
component). That more Ni was needed was clear since the original models were
below the observations at advanced epochs. 

We note that Chugai (2000) and Sollerman et al. (2000) also used 
parameterized models for SN 1998bw somewhat similar to what this 
study finds, but in both cases no consideration was given to the 
implications of an inner dense component. 
Sollerman et al. (2000) used a low velocity zone to reproduce the narrow 
component of the [O I] nebular line. The theoretical need for such a zone, and 
the fact that it should be oxygen-rich, was postulated by Mazzali et al. (2001). 
Chugai (2000) used a dense component to get a good fit to the light curve 
of SN~1998bw also in the context of spherical symmetry, but the 
physical values he derived are very different from those found in 
this study. His model has smaller energy and mass than the original 
hydrodynamical models of Iwamoto et al. (1998) and Nakamura et al. (2001), 
and therefore also smaller than the values derived in this study. 
This is mainly due to the well-known degeneracy in the estimate of mass 
and energy mentioned in Section 3. 
Because when building our model we took into account, at least crudely, 
the results of spectroscopic observations, the values we derived  
are probably more reliable. 

The results of this study help us understand the details of the 
explosion mechanism of hypernovae.
No spherical explosion models so far predicts a high density core near the 
center, but this is required to fit the light curves according to the present 
study.
Actually, recent multi-dimensional models of jet-driven supernova explosions do 
predict such high density material (Khokhlov et al. 1999; Maeda et al. 2002).
Maeda et al. (2002) indeed found that a jet-driven explosion model can explain
the unusual properties of the nebular spectra of SN 1998bw, namely the presence
of iron at higher velocities than oxygen. 

The calculations we have presented, although still in spherical symmetry, take
some of the main features of asymmetric models into account, and therefore they
represent an improvement, although clearly multi-dimensional calculations are
necessary to obtain realistic estimates of the parameters 
(e.g., Equation (4) for illustrative purposes). Our estimate of these 
values (Table 3) based on the spherically symmetric calculations 
presented in section 3 should be regarded as a guide line for future work 
based on aspherical geometry. 
Most of all, our results indirectly support the need for asymmetric explosion 
models. 

H\"oflich et al. (1999) introduced aspherical ejecta to 
reproduce the early phase light curve and polarization of SN 1998bw. 
They derived $E_{51} = 2$, $M_{\rm ej} = 2\msun$, and 
$M(^{56}$Ni) $= 0.2\msun$. 
These values are significantly smaller than hypernova models of Iwamoto 
et al. (1998) and Woosley et al (1999), which were obtained by fitting the 
early phase spectra and light curve (Table 1). 
The caveat is that it is unclear whether the small values of H\"oflich et al. 
(1999) are totally attributed to asphericity. 
They may in fact partly be attributed to the fact that they adopted 
progenitor models of different masses. In fact, using an asymmetric model, 
Maeda et al. (2002) estimated $E_{51} = 10$ based on the late-phase spectra. 
Analyzing a light curve of the early phase 
only yields the possible combination of 
$M_{\rm ej}$ and $E$ (e.g. Arnett 1982; Iwamoto et al. 1998). 
Even in spherical models, it is therefore possible to obtain a good synthetic 
early phase light curve for SN~1998bw using small values 
of $M_{\rm ej}$ and $E$. Such a 
model may however not necessarily reproduce the extremely large velocities 
required by the unusually broad spectral features. 
It thus remains to be examined whether the aspherical model of H\"oflich et al. 
(1999) can reproduce the broad spectral lines in SN 1998bw.

Note also that the outer component determines the light curve near peak, while 
the diffusion time of photons produced in the inner component is too long. 
This implies that a significant fraction of the synthesized $^{56}$Ni must 
reside in the high velocity component. 
This was already required by the original models to reproduce the early rise 
of the light curves, and by the corresponding spectral models to fit the 
strong lines of Fe observed. 
The presence of $^{56}$Ni at high velocity is a clear sign of an asymmetric 
explosion, as noted by Mazzali et al. (2001) and Maeda et al. (2002). 

The accumulation of various studies thus suggests that a jet-driven model 
is very promising to describe hypernova explosions.
In view of claims made for a few normal SNe Ibc, it would also be interesting 
to investigate whether this may not be a common feature of all SNe Ibc, or
maybe even of most, or all, core-collapse SNe. 

From an observational point of view, i.e., spectral evolution and light 
curve shape, the small sample of well-observed type Ic hypernovae can be  
divided into two groups.
One group comprises SNe 1998bw and 2002ap, whose spectra and light curves show 
a similar behavior, while the other group includes SN~1997ef (and possibly 
SN~1997dq, whose spectra and light curve shape are very similar to those 
of SN~1997ef: Matheson et al. 2001). 
We found that the two groups are intrinsically different in their properties, 
as derived in the context of the two-component ejecta model presented here.
The main difference lies in the relative importance of the outer and inner
components, in particular their relative optical depths and the distribution 
of $^{56}$Ni.
The two components of SNe 1998bw and 2002ap have similar relative optical 
depths, but in the case of SN~1997ef the inner component is significantly 
more prominent: the ratios $\tau_{\rm in}/\tau_{\rm out}$ (or equivalently 
$M_{\rm in}/M_{\rm out}$) and $M$($^{56}$Ni)$_{\rm in}/M$($^{56}$Ni)$_{\rm out}$ 
are similar in SNe 1998bw and 2002ap, while they are much larger in SN 1997ef.
This is also supported by the late-phase spectra: both SNe 1998bw and 2002ap 
entered the fully nebular phase somewhere between day 100 and 150, while 
SN~1997ef (and SN~1997dq) at a comparable epoch still preserved a photospheric 
component with very small velocities (P. Mazzali et al., in preparation).

In the context of a jet-driven model, the difference in the relative
importance of the outer and inner components could be ascribed to the 
strength of the jets.
The jet in SN~1997ef could be less efficient in turning its energy into the
explosion of the entire star, thus accretion onto the center might continue
for a longer period of time than in SNe 1998bw and 2002ap.
This argument is strengthened by the other finding of this study, namely
that the jetted matter was blown up and mixed into the outer ejecta more  
effectively in SNe 1998bw and 2002ap than in SN~1997ef, as shown by the 
larger fraction of $^{56}$Ni present in the outer regions of the former 
two hypernovae.

It is also very likely that the observed light curve of a complex event 
such as a hypernova explosion is different depending on the line of sight, 
especially at early times. 
H\"oflich et al. (1999) found that the peak luminosity of an 
oblate ellipsoidal SN, which is produced by a jet-induced explosion 
(Khokhlov et al. 1999: Maeda et al. 2002) is smaller for a larger angle. 
The relatively small estimated value of $M$($^{56}$Ni)$_{\rm out}$ in 
SN~1997ef may thus be partly ascribed to SN~1997ef being viewed from off the 
jet axis. 

The spectroscopic properties of our two-component models should be examined in
detail, beyond the crude approximation used in this work, and compared with
observations at both early and late phases, to verify that they are indeed
consistent.
Also, hydrodynamical models which satisfy the conditions found in this study,
e.g. the existence of a dense core, should be investigated, as should the
effects of multi dimensionality in the radiation transport in SN ejecta. 
These issues have been partially addressed by various authors 
(hydrodynamical models: Khokhlov et al. 1999: 
hydrodunamical and nucleosynthetic models: Maeda \& Nomoto 2003a, 2003b), 
their results are at least qualitatively in good agreement 
with those of this light curve study.

\acknowledgments
This work has been supported in part by the 
grant-in-Aid for Scientific Research (12640233, 14047206, 14540223),  
and COE research (07CE2002) of the Ministry of Education, 
Science, Culture, Sports, and Technology in Japan.

\clearpage




\begin{figure*}
\begin{center}
\epsscale{1.7}

\plotone{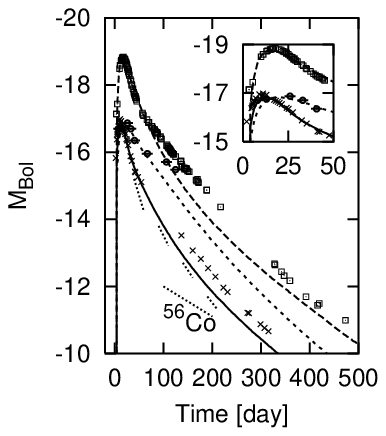}
\caption{Light curves based on the original 
hydrodynamical models (Table 1)
for SNe 2002ap (solid), 1998bw (long-dashed) and 1997ef (short-dashed).
Also shown are the observational bolometric points of SNe 2002ap (crosses:
constructed from MAGNUM (Yoshii et al. 2003; Y. Yoshii et al., in preparation), 
Wise Observatory (Gal-Yam et al. 2002), 
and State Observatory, India (Pandey ey al. 2003)), 
1998bw (squares: taken from Patat et al. 2001), and 
1997ef (circles: Mazzali et al. 2000). 
The dotted lines show the decline rates at 50, 100, 150, and 200 days predicted 
by equation (6), as well as the full trapping line, corresponding to the 
$^{56}$Co (denoted with $^{56}$Co).
\label{fig1}
}
\end{center}

\end{figure*}

\clearpage

\begin{figure*}
\begin{center}
\epsscale{1.7}

\plotone{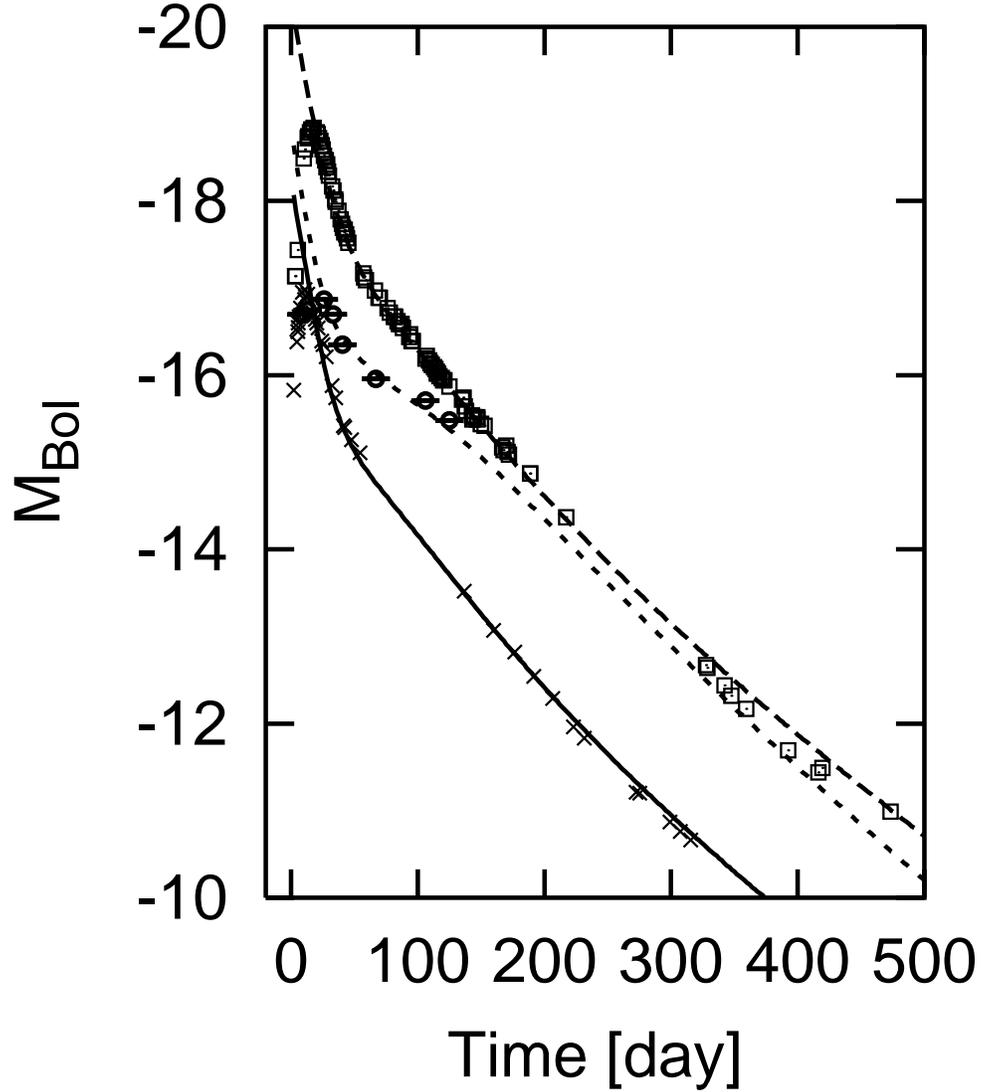}
\caption{(a)The results of the simple $\gamma$-ray deposition 
computation for the
two-component models based on equation (7) (for the meaning of the symbols,
see the caption of Figure 1).
The parameters used to fit the observed points are listed in Table 2.
(b-d) For the individual objects, the contributions from each component are
shown by dotted lines (the outer component) and by dashed lines (the inner component),
while the total luminosity is shown as solid lines.
For each component, the thin line is the luminosity emitted as $\gamma$-rays and positrons,
and the thick line is the luminosity deposited in the ejecta (i.e., $L_{\rm opt}$).
\label{fig2}
}
\end{center}

\end{figure*}

\clearpage

\begin{figure*}
\begin{center}
\epsscale{1.7}

\plotone{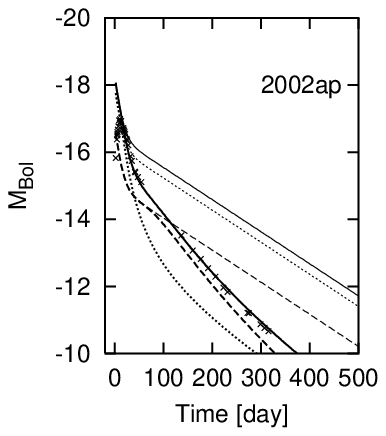}
\end{center}
\end{figure*}

\clearpage

\begin{figure*}
\begin{center}
\epsscale{1.7}

\plotone{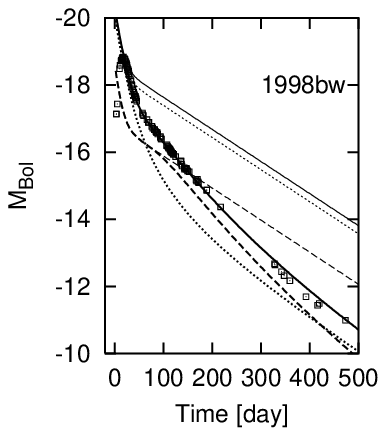}
\end{center}

\end{figure*}

\clearpage

\begin{figure*}
\begin{center}
\epsscale{1.7}

\plotone{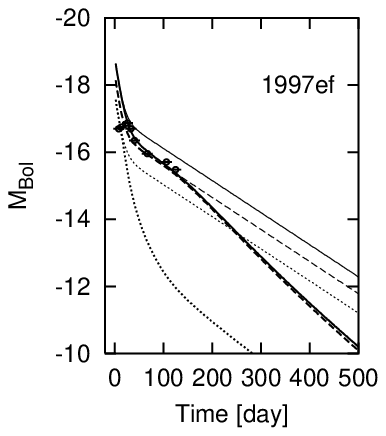}
\end{center}

\end{figure*}

\clearpage

\begin{figure*}
\begin{center}
\epsscale{1.7}

\plotone{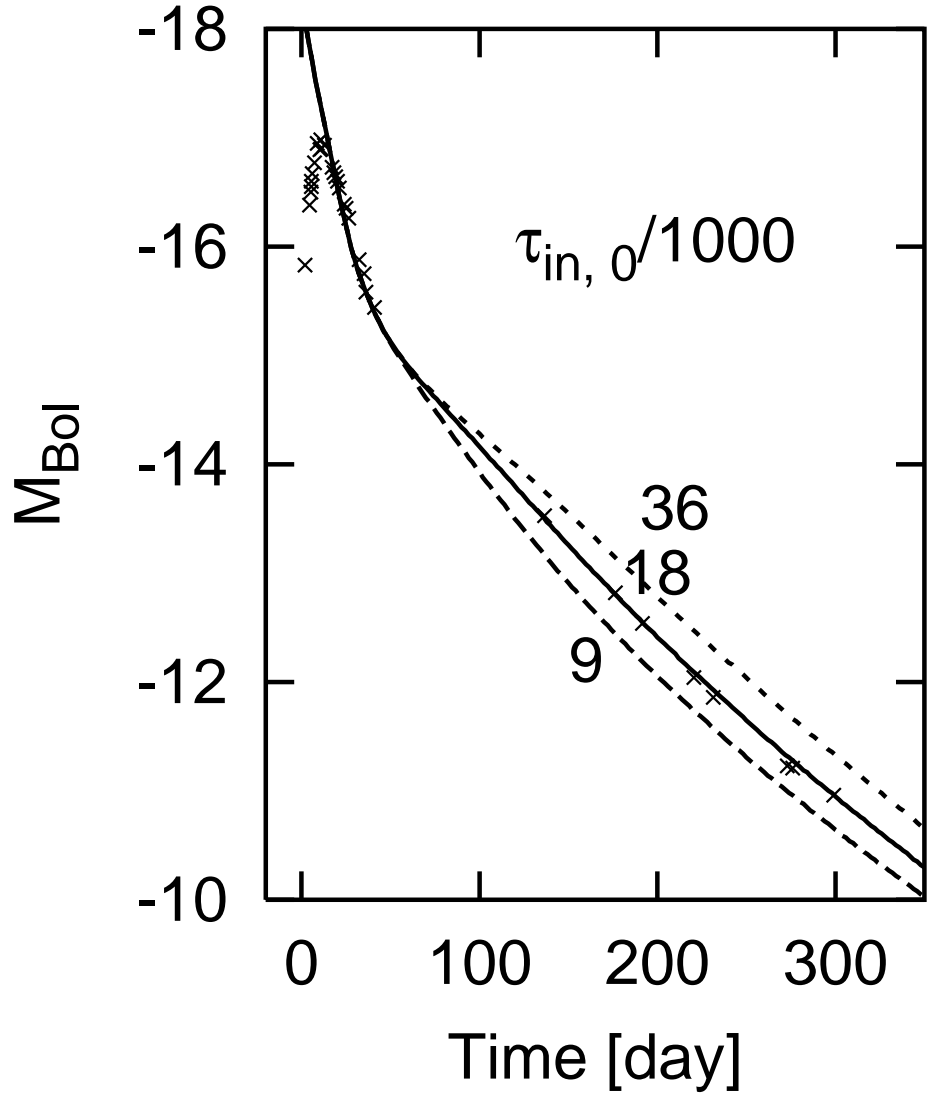}
\caption{Dependence on 
(a) $\tau_{{\rm in},0}$, (b) 
$\tau_{{\rm out},0}$, and (c) 
$M$($^{56}$Ni)$_{\rm in}$/$M$($^{56}$Ni)$_{\rm out}$. 
If not explicitly mentioned, the values of parameters are set to 
$\tau_{{\rm in},0}/1000 = 18$, $\tau_{{\rm out},0}/1000 = 0.6$, 
$M$($^{56}$Ni)$_{\rm in}$/$M$($^{56}$Ni)$_{\rm out} = 0.33$, and 
$M$($^{56}$Ni)$_{\rm tot} = 0.08\msun$. 
For comparison, the observational bolometric points of SN 2002ap 
are shown. 
\label{fig3}
}
\end{center}

\end{figure*}

\clearpage

\begin{figure*}
\begin{center}
\epsscale{1.7}

\plotone{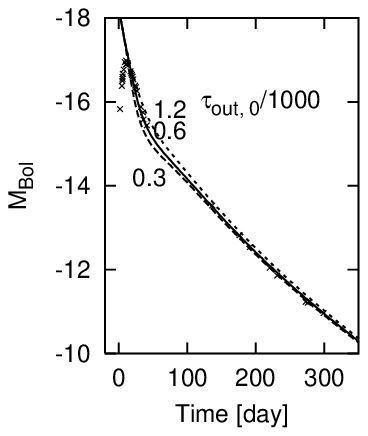}
\end{center}

\end{figure*}

\clearpage

\begin{figure*}
\begin{center}
\epsscale{1.7}

\plotone{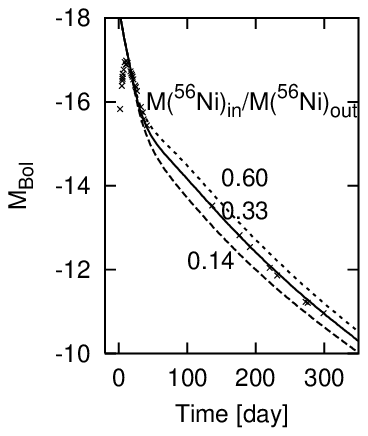}
\end{center}

\end{figure*}

\clearpage

\begin{figure*}
\begin{center}
\epsscale{1.7}

\plotone{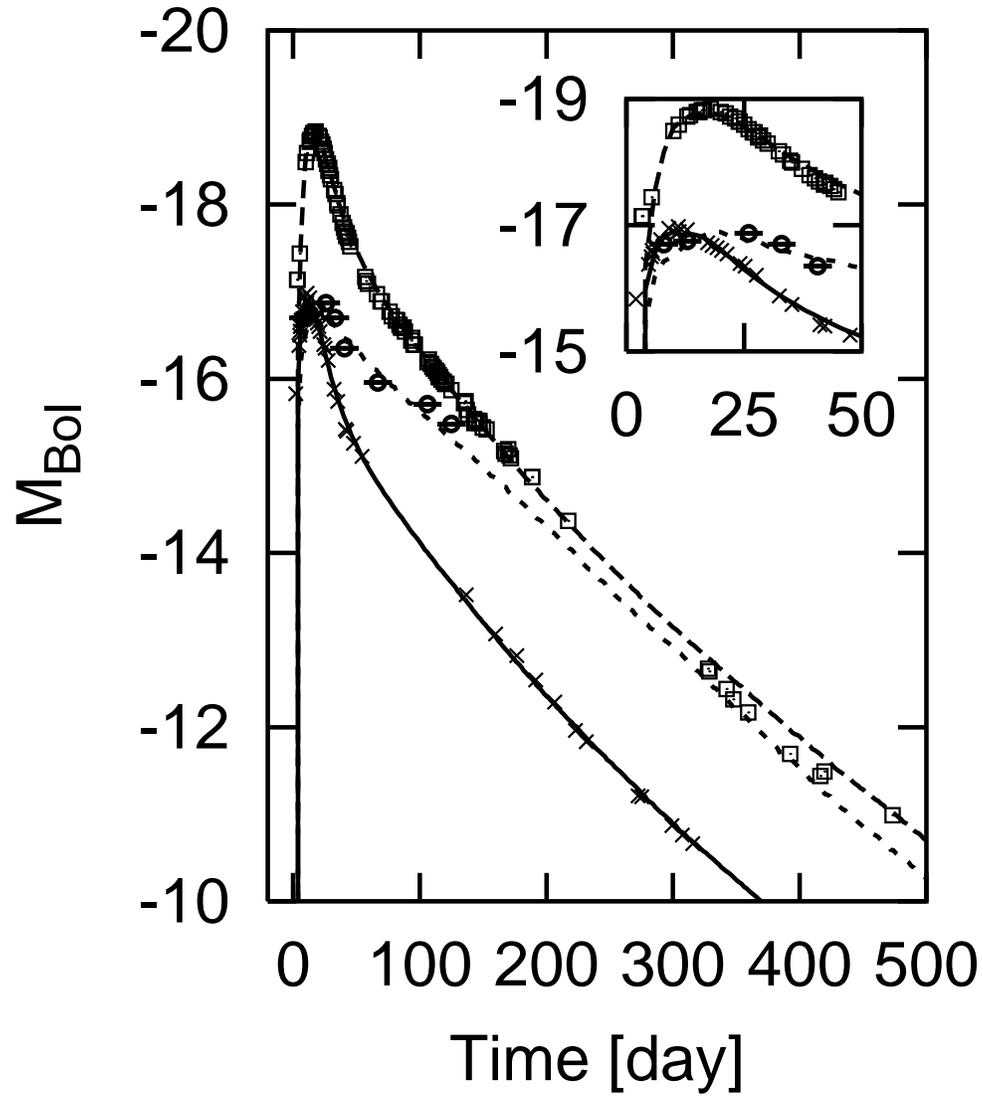}
\caption{Synthetic light curves of the two-component models (Table 3),
computed with the Monte Carlo light curve code described in Section 3.
The meaning of the symbols is the same as in Figure 1.
\label{fig4}
}
\end{center}

\end{figure*}

\clearpage



\begin{deluxetable}{cccccccc}
\rotate

\tabletypesize{\scriptsize}
\tablecaption{The original models used 
to fit the early light curves\tablenotemark{a}.}
\tablewidth{0pt}
\tablehead{
\colhead{SN} &
\colhead{Model} &
\colhead{$M_{\rm ej}$ ($\msun$)}   & 
\colhead{$E_{51}$} &
\colhead{$M$($^{56}$Ni) ($\msun$)} & 
\colhead{$\tau_{0}/1000$\tablenotemark{b}}  &
\colhead{$t_{\gamma}$\tablenotemark{b}} &
\colhead{Declining rate\tablenotemark{c}}
}
\startdata
1997ef & CO100   & 9.5  & 21  & 0.11 & 4.3 &  66 & 0.010\\
1998bw & CO138     & 10.2 & 45  & 0.5  & 2.3 &  48 & 0.018\\
2002ap & CO100resc & 2.4  & 5.4 & 0.07 & 1   &  33 & 0.018\\
\enddata
\tablenotetext{a}{Taken from Mazzali et al. (2002) for SN~2002ap, 
Nakamura et al. (2001) for SN~1998bw, 
and Mazzali et al. (2000) for SN~1997ef.}
\tablenotetext{b}{Computed based on the original models. See section 2
for the detail.}
\tablenotetext{c}{The decline rates (mags day$^{-1}$) observed at days $50 - 200$.}
\end{deluxetable}

\clearpage

\begin{deluxetable}{ccccccll}
\rotate
\tabletypesize{\scriptsize}
\tablecaption{The two-component models for the simple deposition computation.}
\tablewidth{0pt}
\tablehead{
\colhead{} &
\colhead{$\tau_{{\rm in},0}/1000$}  & 
\colhead{$M(^{56}$Ni)$_{\rm in}$ ($\msun$)}&
\colhead{$\tau_{{\rm out},0}/1000$} & 
\colhead{$M(^{56}$Ni)$_{\rm out}$ ($\msun$)}  &
\colhead{$\tau_{\rm in}/\tau_{\rm out}$} &
\colhead{$M(^{56}$Ni)$_{\rm in}$/$M$($^{56}$Ni)$_{\rm out}$}&
\colhead{$M(^{56}$Ni)$_{\rm tot}$ ($\msun$)}
}
\startdata
1997ef & 50 & 0.085 & 0.6 & 0.05 & 83 & 1.70 & 0.135 \\
1998bw & 26 & 0.11  & 1   & 0.44 & 26 & 0.25 & 0.55 \\
2002ap & 18 & 0.02  & 0.6 & 0.06 & 30 & 0.33 & 0.08 \\
\enddata
\end{deluxetable}

\clearpage

\begin{deluxetable}{ccccccccccc}
\rotate
\tabletypesize{\scriptsize}
\tablecaption{The two-component light curve models.}
\tablewidth{0pt}
\tablehead{
\colhead{SN} &
\colhead{$v_{\rm in}$ (km s$^{-1}$)} & 

\colhead{$M_{\rm in}$ ($\msun$)} &
\colhead{$M$($^{56}$Ni)$_{\rm in}$ ($\msun$)}& 

\colhead{$E_{{\rm in},51}$}  &
\colhead{$M_{\rm out}$ ($\msun$)} &
\colhead{$M$($^{56}$Ni)$_{\rm out}$ ($\msun$)} &
\colhead{$E_{{\rm out},51}$} &
\colhead{$M_{\rm tot}$ ($\msun$)} &
\colhead{$M$($^{56}$Ni)$_{\rm tot}$ ($\msun$)}& 
\colhead{$E_{{\rm tot},51}$} 
}
\startdata
1997ef & 3500 & 5.0 & 0.084 & 0.37 & 2.9  & 0.050 & 21  &  7.9 & 0.134 & 21.4 \\
1998bw & 5000 & 3.9 & 0.116 & 0.58 & 6.8  & 0.435 & 45  & 10.7 & 0.551 & 45.6 \\
2002ap & 3000 & 1.1 & 0.014 & 0.06 & 1.6  & 0.065 & 5.4 &  2.7 & 0.079 &  5.5 \\
\enddata
\end{deluxetable}

\end{document}